\documentclass[pre,twocolumn,aps,showpacs]{revtex4-1}

\usepackage{amsmath}
\usepackage{amssymb}
\usepackage{graphicx}
\usepackage{color}

\begin{document}
 
\title{Fluctuations in partitioning systems with few degrees of freedom}

\author{L. Cerino} 
\affiliation{Dipartimento di Fisica, Universit\`a di
  Roma Sapienza, p.le A. Moro 2, 00185 Roma, Italy}
\author{G. Gradenigo} 
\affiliation{CEA/DSM-CNRS/URA 2306, CEA Saclay,
  F-91191 Gif-sur-Yvette Cedex, France} 
\affiliation{Univ. Paris Sud;
  CNRS; LPTMS, UMR 8626, Orsay 91405 France} 
\author{A. Sarracino}
\affiliation{ISC-CNR and Dipartimento di Fisica, Universit\`a Sapienza,
  p.le A. Moro 2, 00185 Roma, Italy,} 
\affiliation{Kavli Institute for
  Theoretical Physics China, CAS, Beijing 100190, China} 
\affiliation{
  Laboratoire de Physique Th\'eorique de la Mati\`ere Condens\'ee,
  CNRS UMR 7600, case courrier 121, Universit\'e Paris 6, 4 Place
  Jussieu, 75255 Paris Cedex} 
\author{D. Villamaina}
\affiliation{Laboratoire de Physique Th\'eorique de l'ENS and Institut de
  Physique Th\'eorique Philippe Meyer, 24 rue Lhomond 75005 Paris -
  France} 
\author{A. Vulpiani} 
\affiliation{Dipartimento di Fisica,
  Universit\`a Sapienza and ISC-CNR, p.le A. Moro 2, 00185 Roma,
  Italy} 
\affiliation{Kavli Institute for Theoretical Physics China,
  CAS, Beijing 100190, China}

\begin{abstract}
  We study the behavior of a moving wall in contact with a particle
  gas and subjected to an external force.  We compare the fluctuations
  of the system observed in the microcanonical and canonical
  ensembles, at varying the number of particles. Static and dynamic
  correlations signal significant differences between the two
  ensembles. Furthermore, velocity-velocity correlations of the moving
  wall present a complex two-time relaxation which cannot be
  reproduced by a standard Langevin-like description. Quite
  remarkably, increasing the number of gas particles in an elongated
  geometry, we find a typical timescale, related to the interaction
  between the partitioning wall and the particles, which grows
  macroscopically.
\end{abstract}


\maketitle

\section{Introduction}

Macroscopic objects contain, at least, $N=O(10^{20})$ particles,
therefore in the mathematical modeling, one can safely assume $N \to
\infty$ and study their asymptotic features (e.g. the thermodynamics
limit).  As consequence of such a huge value of $N$, up to few decades
ago statistical mechanics has been devoted almost only to the study of
systems with many degrees of freedom\cite{CFLV08}. On the other hand the present
day instrumentation allows us for the manipulation (and sometimes
control) of small systems at micro, and even nano, scales; it is not
necessary to emphasize the practical relevance of small
systems~\cite{BLR05,SC11}.

In order to deal with systems with a small number of particles, say
$O(10^2)$ or less, we are forced to (re)consider in details some
aspects of the statistical mechanics~\cite{LPV67} which for
macroscopic bodies are not very relevant.  For instance in large
systems the fluctuations are always relatively negligible (and
apparently) irrelevant~\cite{review}.  In a similar way, for
macroscopic objects, there are not particular problems for the
definition of temperature~\cite{FPSVV11} neither significant
differences using different statistical ensembles (e.g.
microcanonical or canonical).

Among the physical systems relevant for the nanosciences we can
mention the class of partitioning objects containing an extra degree
of freedom (a wall) which separates the system into subsystems.  A
paradigmatic example is given by the adiabatic
piston~\cite{CPS96,GP99,GPL03,BKM04,CPPV07}: a system of $N$ particles
of mass $m$ (e.g. an ideal gas) in a container of length $L$ and
cross-section $A$, separated in two regions by a movable wall (the
piston) of mass $M$.  The walls of the container are supposed to be
perfect insulators preventing any mass or heat exchanges with the
exterior.  Gas particles undergo purely elastic collisions with the
piston and the walls, and the piston is constrained to move along one
axis.  If at initial time the temperatures $T_L,\, T_R$ and pressures
$P_L,\, P_R$ in the left and right parts do not coincide, the system
shows a rather rich phenomenology (depending on $M/m$, $N/L$ etc) in
the approach to the mechanical and thermodynamic equilibrium.

A physical version of the adiabatic piston is a big Brownian particle
sliding along a microtubule filled with particles~\cite{delre}.  The
authors of ref~\cite{delre} showed how the presence of the wall is
able to induce, even in the equilibrium state, rather complex (and
slow) dynamical behavior.

Our paper is devoted to the statistical mechanics of a system similar
to a piston where particles are confined in a tube with a non fixed
wall, on which an external force acts, see Fig.~\ref{fig1}. The
pressure on the piston due to the interaction with the gas particles
on one side is balanced by the external force, so that the piston
reaches a stationary state. We are interested in the study of piston
fluctuations (of position and velocity) around the equilibrium
state. In the case of non interacting particles it is possible to find
in an exact way the equilibrium properties of the system both in
microcanonical and canonical ensembles (this latter case is realized
by putting a thermostat on the fixed wall, which thermalizes particles
colliding with the wall).  One obtains that, even in the limit $N\gg
1$, the fluctuations of the wall position are different in the
canonical and microcanonical ensembles. As important consequence of
such a difference, which holds also for the interacting particles, we
have that the correlation function (of the velocity) $C(t)$ must be
different in the two ensembles.

Numerical simulations show a non trivial behavior of $C(t)$ with a
negative minimum around a characteristic time $\tau(N)$ increasing
linearly with $N$.  A comparison between the numerical results and an
appropriate Langevin equation shows how even for large $N$ the
presence of the wall has non trivial consequences which can have a
role for an effective modeling of the system.

The paper is organized as follows: Section 2 describes the model in
detail and presents the analytical results for the ideal gas case; in
Section 3 we report the results of molecular dynamics simulations in
the interacting case. Section 4 is devoted to the derivation of an
effective Langevin equation for describing the dynamics of the piston,
and, finally, in Section 5 some conclusions are drawn. Two Appendices
provide details about the computations.

\section{The model}

\begin{figure}[!t]
\includegraphics[width=.8\columnwidth,clip=true]{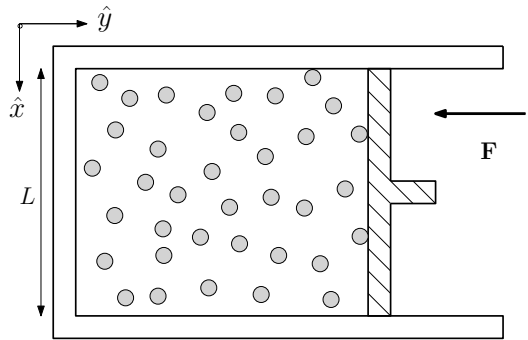}
\caption{Sketch of the piston model: a gas of particles is confined by
  a moving wall which is subjected to a constant external force.}
\label{fig1}
\end{figure}

We consider a two-dimensional system composed by a gas of $N$
point-like particles with mass $m$, positions
$\mathbf{x}_i=\{x_i,y_i\}$ and momentum $\mathbf{p}_i$, with
$i=1,\ldots,N$, contained in a rectangular box with one moving
adiabatic wall of length $L$ (hereafter referred to as the
``piston''). The position of the piston is denoted by $Y$ and its
momentum and mass are $P$ and $M$, respectively (see Fig~\ref{fig1}
for visual explanation). An external force
$\mathbf{F}=-F\cdot\hat{y}$, directed along the horizontal axis
$\hat{y}$, acts on the piston, which is also subject to the collisions
with the particles.  In the tubular geometry that we consider,
in which the size of the sistem is increased anisotropically only
  along one direction when adding particles, the
piston plays the role of a ``partitioning'' object with respect to the
particle gas, namely its position determines the volume available for
the gas. This system has been studied in~\cite{FPSVV11} as an
effective thermometer model. In the following the particle-particle
and particle-piston interactions are described in a Hamiltonian
(conservative) context and the piston can slide without dissipation
along the $y$ axis. The case of dissipative interactions, inducing
nonequilibrium behaviors, of similar systems have been studied for
instance in~\cite{BRB05,HS06,CMP08,FKS12,SGP13,SH14}.

%\subsection{Analytical results for non-interacting gas}

We start by considering the case of a non-interacting gas, so that the
Hamiltonian of the system reads
\begin{equation}\label{Hamiltoniana}
\mathcal{H}=\sum_{i=1}^{N}\frac{|\mathbf{p}_i|^2}{2m} + \frac{P^2}{2M} +FY,
\end{equation}
with geometrical constraints 
\begin{equation}\label{vincoli}
\left\{
\begin{array}{l}
Y>0;\\
0<x_i<L;\\
0<y_i<Y.\\
\end{array}
\right.
\end{equation}

We are interested in the study of the behavior of fluctuations at
varying the number of gas particles, and, in particular, in the
comparison between the microcanonical and canonical ensembles.  As
shown in the Appendix A, in the microcanonical ensemble the
temperature of the system is related to the energy $E$ of the system
by the relation
\begin{equation}\label{temperature_micro}
k_BT=k^2_B\left(\frac{\partial \log \Sigma(E)}{\partial E}\right)^{-1}=\frac{E}{2N+\frac{3}{2}},
\end{equation}
where 
\begin{equation}
\Sigma(E) = \int_{\mathcal{H}<E}d^N\mathbf{x}~d^N\mathbf{p}~dY~dP
\end{equation}
is the phase space volume and $k_B$ the Boltzmann's constant. The
static properties of this system, average position $\langle Y \rangle$
and variance $\sigma_Y^2=\langle Y^2 \rangle$, can be readily obtained
(see Appendix A), yielding
\begin{eqnarray}\label{sm}
\langle Y \rangle &=& \frac{(N + 1)k_BT}{F}, \\
\sigma_Y^2&=& \frac{(N + 1/2)(N+1)}{2N+5/2}\left(\frac{k_BT}{F}\right)^2.\label{sm2}
\end{eqnarray}

Let us open a parenthesis on the definition of
  temperature. Eq.~(\ref{temperature_micro}) is not the unique
  possibility, another way is via the formula
\begin{equation}
k_B T'=k_B^2\left(\frac{\partial \ln \omega(E)}{\partial E}\right)^{-1}
\end{equation}
where $\omega (E) =\frac{\partial \Sigma (E)}{\partial E}$. There are
cases where $T$ and $T'$ can be different and, in particular, $T'$ can
be negative, e.g. in the case of point vortex systems~\cite{DS14}. On
the other hand, in a perfect gas it is easy to see that the two
definitions are equivalent for $N\gg 1$ since $T-T' =
O\left(\frac{1}{N}\right)$~\cite{HUANG}; this result also holds for
weakly interacting systems.

Analogous results can be obtained for the canonical case, where the
system is in contact with a reservoir at temperature $T$. In this
case, the energy of the system is
\begin{equation}\label{temperatura_can}
E=-\frac{\partial \log Z(\beta)}{\partial \beta}=\left(2N+\frac{2}{3}\right)k_B T,
\end{equation}
where 
\begin{equation}
Z(\beta) = \int d^N\mathbf{x}~d^N\mathbf{p}~dY~dP~e^{-\beta\mathcal{H}}
\end{equation}
and $\beta=1/(k_BT)$ is the inverse temperature.  Average position and
variance $\sigma_Y^2$ read (see Appendix A)
\begin{eqnarray}\label{sc}
\langle Y \rangle &=& \frac{(N + 1)k_BT}{F}, \\
\sigma_Y^2&=& (N+1)\left(\frac{k_BT}{F}\right)^2.\label{sc2}
\end{eqnarray}

%\subsection{Comparison between microcanonical and canonical ensemble}

In order to compare the results for static quantities $\langle Y
\rangle$ and $\sigma_Y^2$ in the two ensembles, for each temperature $T$
in the canonical ensemble we consider the corresponding energy in the
microcanonical, such that $T=E/(2Nk_B)$, in the limit of large number
of particles $N\gg 1$.  While the average position is always the same,
from Eqs.~(\ref{sm2}) and~(\ref{sc2}) one observes that fluctuations
differ by a factor $1/2$, also in the large $N$ limit. In the
Table~\ref{table} we summarize these findings.

The equivalence of ensembles in the thermodynamic limit is expected
only for average values, and not for
fluctuations~\cite{LPV67,PM96}. Indeed, the observed ``discrepancy''
is explained by noting that the variance in the canonical ensemble can
be expressed as the sum of two contributions, namely a term which
corresponds to the variance of the piston in the microcanonical
ensemble at fixed energy plus a term corresponding to energy
fluctuations at fixed temperature:
\begin{equation}\label{somma}
\sigma_Y^2(T)= \left .\alpha \sigma_E^2(T)+\sigma_Y^2(E)\right|_{E=\langle E\rangle_\beta},
\end{equation}
where $\alpha=1/(4F^2)+O(1/N)$ and
$\sigma_E^2=\langle\mathcal{H}^2\rangle - E^2$. Therefore, for $N\gg
1$, since $\sigma_E^2(T)\approx 2N(k_BT)^2$, one has
$\sigma_Y^2(T)=2\sigma_Y^2(E)|_{E=2Nk_BT}$.

Let us open a short digression on terminology.  With the term
``canonical ensemble'' we mean the system with Hamiltonian in
Eq.~(\ref{Hamiltoniana}) and Eq.~(\ref{vincoli}) (in the following we
will include also the interactions among the particles) interacting
with a thermal bath at temperature $T$. Noting that the pressure is
nothing but $F/L$, one can then say that we are dealing with an
ensemble at fixed temperature and fixed pressure for the system
without the terms $FX$ and $P^2/(2M)$ in the
Hamiltonian~\cite{PM96}. In a similar way our microcanonical ensemble
correspond to an ensemble with fixed enthalpy for the system without
the terms $FX$ and $P^2/(2M)$ in the Hamiltonian. We prefer the terms
canonical and microcanonical because they put the dynamical variables
describing the wall on the same level of those for the particles. Let
us note that the mass of the piston is important for the dynamical
properties.

\begin{table}
\centering
\begin{tabular}{c|c|c}
&\textbf{Canonical}&\textbf{Microcanonical} \\ 
\hline Temperature: & $T$ & $\frac{E}{2Nk_B}$ \\ 
$\langle Y \rangle$ & $\frac{Nk_BT}{F}$ & $\frac{E}{2F}=\frac{Nk_BT}{F}$ \\ 
$\sigma^2_Y$&$\frac{N(k_BT)^2}{F^2}$&$\frac{E^2}{8NF^2}=\frac{N(k_BT)^2}{2F^2}$ \\
\end{tabular}
\caption{Comparison of average position and variance in the
  microcanonical and canonical ensembles.}
\label{table} 
\end{table}

The above results on the fluctuations immediately produce two
important consequences on the dynamical correlations in the two
ensembles. First, notice that the finite value of the variance
$\sigma_Y^2$ in both cases for finite $N$ implies that the diffusion
coefficient $D$ of the piston is zero, implying that the piston
remains confined.  Second, the difference in the static fluctuations
have repercussions on the shape of the velocity-velocity fluctuations
in the canonical and microcanonical ensemble.  Let us note that
\begin{equation}
\sigma_Y^2=\langle (Y-\langle Y\rangle)^2\rangle=\int_0^\infty\int_0^\infty\langle V(t')V(t'')\rangle dt' dt'',
\end{equation}
where $V(t)$ is the velocity of the piston. Since $\sigma_Y^2$ are
different in the canonical and microcanonical ensembles also the
correlation $\langle V(t)V(0)\rangle$ must be different.  These issues
will be addressed in the next section, in the case of interacting gas.

Exactly the same considerations about the difference of fluctuations
in the canonical and microcanonical ensembles hold in the case that a
different termodynamic limit is considered, in wich the size of the
piston is increased isotropically. In this case, in order to have that
for each value of $N$ the shape of the gas compartement is isotropic,
namely $\langle Y \rangle = L$, and that the density $\rho = N /L^2$
and the pressure $p=F/L$ are constant, we need the scaling $F \sim
\sqrt{N}$ for the force acting on the piston. If we insert such
scaling for $F$ in the equations Eq.~(\ref{sm2},\ref{sc2}), we find
that increasing isotropically the size of the compartiment, at
variance with the tubular geometry, the mean square dispacement
$\sigma_Y^2$ of the partitioning wall becomes asimptotically costant
for increasing $N$ in the two ensembles. On the contrary the factor
$2$ by which canonical and microcanical fluctuations differ remains
the same. The comparison between the two different thermodynamic
limits tell us on one hand that the result on the difference in
canonical and microcanonical fluctuations is robust and on the other
hand allows us to point out the peculiarities of the tubular geometry.

\section{Numerical simulations for the interacting case}

In order to understand whether the previous results are peculiar to
the non-interacting case, and to study a more realistic case, we
perform molecular dynamics simulations of the system with an
interacting particle gas. We consider a repulsive interaction
potential $V(\mathbf{r})$ for soft disks, with cut-off $r_c$
\begin{equation}\label{potenziale riscalato}
V(\mathbf{r})=
\left\{
\begin{array}{l}
V_0\Big[\left(\frac{r_0}{r}\right)^{12} - \left(\frac{r_0}{r_c}\right)^{12} 
+ 12\left(\frac{r_0}{r_c}\right)^{12} \left(\frac{r}{r_c}-1\right)\Big]\\
\qquad\qquad\qquad\qquad\qquad\qquad\qquad \textrm{for } r<r_c \\
0 \qquad \textrm{for } r>r_c,
\end{array}
\right.
\end{equation}
where $r=|\mathbf{r}|$ is the distance between particles, $V_0$ is the
potential intensity and $r_0$ is the average interaction range.  The
same potential also describes the interaction of particles with
walls. In the simulations of the canonical ensemble the coupling with
the \emph{reservoir} at temperature $T$ is implemented in the
following way. We consider that the side of the box opposite to the
piston acts as a thermostat, so that when a particle enters the
interaction region with the wall, namely its distance from the wall is
smaller than $r_0$, the velocity is changed along the $y$ axis
according to the Maxwellian distribution $p(v_y)\propto v_y
\exp(-v_y^2/2mk_BT)$, for $v_y>0$~\cite{TTKB98}.  The study of the
system upon varying $N$ is performed by retaining a tubular geometry,
namely keeping the length $L$ and the force $F$ constant and letting
the equilibrium position $\langle Y\rangle$ increase accordingly, so
that the gas density remains fixed. The results here described are not
related to a specific interacion. Ideed, we also studied the case of a
stronger interaction potential $V(r)\sim r^{-64}$, which at low
density reproduces the behavior of hard-disk statistics~\cite{VT14},
finding analogous results.

%\subsection{Statics}

\begin{figure}[!t]
\includegraphics[width=.8\columnwidth,clip=true]{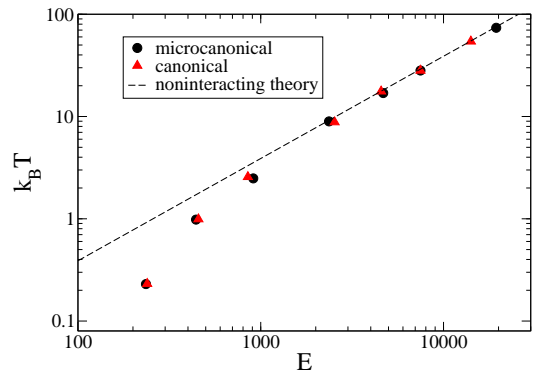}
\caption{(color online) The temperature $k_BT$ (in the microcanonical ensemble is $M\left<V^{2}\right>$) is plotted as a function of energy $E$
  (in the canonical ensemble $E=\langle \mathcal{H}\rangle$) for
  $N=128$. The dashed line represents the theoretical result for
  non-interacting particles $k_BT=E/(2N+3/2)$, which is expected to
  hold for high temperatures. Other parameters in the simulations are
  $L=10$, $F=10$, $m=1$ and $M=128$.}
\label{fig2}
\end{figure}

\begin{figure}[!t]
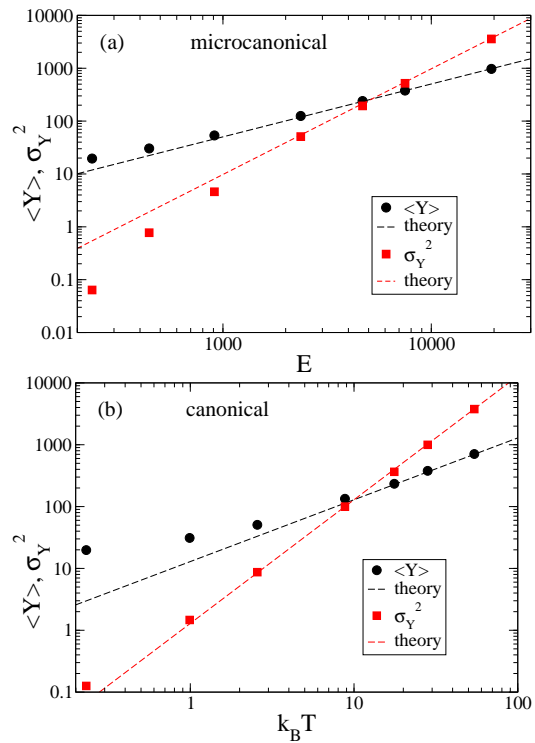

\includegraphics[width=.8\columnwidth,clip=true]{Y_E_micro.eps}
\includegraphics[width=.8\columnwidth,clip=true]{Y_E_can.eps}
\caption{(color online) Panel (a): The average position $\langle Y\rangle$ and variance
  $\sigma_Y^2$ are plotted as a function of energy $E$ in the
  microcanonical ensemble with $N=128$.  Dashed lines represent the
  theoretical results for the non-interacting gas: $\langle
  Y\rangle=(N+1)E/(2N+3/2)$ and
  $\sigma_Y^2=(N+1)(N+1/2)/[(2N+5/2)(2N+3/2)^2](E/F)^2$. Panel (b): Same
  quantities as a function of $k_BT$ in the canonical
  ensemble. Theoretical results for the non-interacting gas are:
  $\langle Y\rangle=(N+1)k_BT/F)$ and $\sigma_Y^2=(N+1)(k_BT)^2/F^2$.
  Other parameters in the simulations are $L=10$, $F=10$, $m=1$ and
  $M=128$.}
\label{fig3}
\end{figure}

We start the numerical study of this interacting case by checking the
validity of the relation~(\ref{temperature_micro}). In
Figure~\ref{fig2} we plot the temperature $T$ as a function of the
energy $E$ in the microcanonical and canonical ensembles. The temperature is computed as $k_B T = M\left<V^{2}\right>$ whereas energy is $E=\langle
\mathcal{H}\rangle$. As expected, the theoretical
relation~(\ref{temperature_micro}) derived in the non-interacting
system is valid at high temperatures, where interactions become
negligible. In Figure~\ref{fig3} we report the average values of the
piston position and its variance in the two ensembles. Notice that
also in this case the analytical predictions~(\ref{sm}) and~(\ref{sc})
hold in the high energy (or temperature) regions.

\begin{figure}[!t]
\includegraphics[width=.8\columnwidth,clip=true]{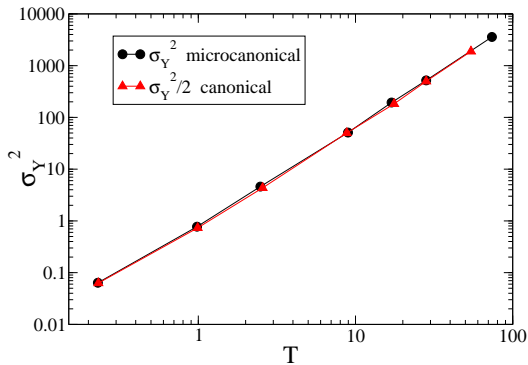}
\caption{(color online) The variance $\sigma_Y^2$ is plotted as a function of $k_BT$
  in the canonical and microcanonical ensembles (in the latter case we
  consider simulations at constant energy and the temperature is
  obtained from $k_BT=M\langle V^2\rangle_E$) for $N=128$. Other
  parameters in the simulations are $L=10$, $F=10$, $m=1$ and
  $M=128$.}
\label{fig3bis}
\end{figure}

It is interesting the fact that also in the interacting case the
factor $1/2$ between the $\sigma_Y^2$ in the canonical and
microcanonical is still present (see Fig.~\ref{fig3bis}).

%\subsection{Dynamics}

\begin{figure}[!t]
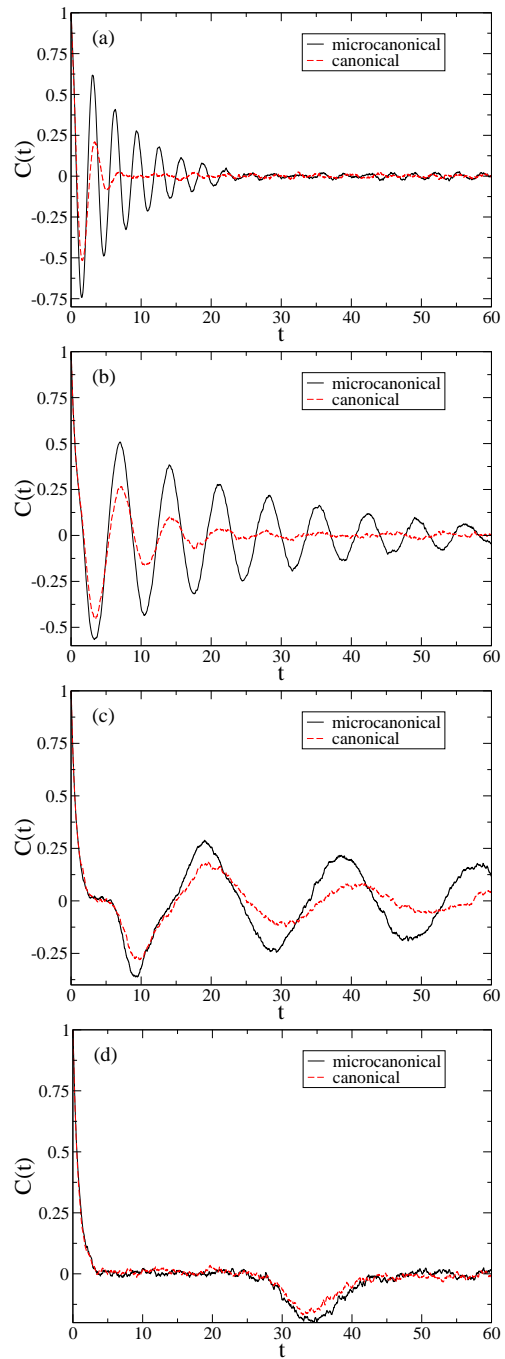

\includegraphics[width=.75\columnwidth,clip=true]{corrN16.eps}
\includegraphics[width=.75\columnwidth,clip=true]{corrN64.eps}
\includegraphics[width=.75\columnwidth,clip=true]{corrN256.eps}
\includegraphics[width=.75\columnwidth,clip=true]{corrN1024.eps}
\caption{(color online) Velocity autocorrelation functions of the piston in the
  microcanonical and canonical ensembles, for $N=16$, panel (a), $N=64$, panel (b),
$N=256$, panel (c)
  and $N=1024$, panel (d). Other parameters are $L=30$,
  $F=150$, $T=10$, $m=1$ and $M=50$.}
\label{fig4}
\end{figure}

Interesting behaviors are also found for the dynamical properties of
this system. Indeed, differences in the fluctuations between
microcanonical and canonical are evident from the study of correlation
functions. In particular, in Figure~\ref{fig4} we compare the behavior
of the normalized velocity autocorrelation function of the piston,
$C(t)=\langle V(t)V(0)\rangle/\langle V(0)V(0)\rangle$, for different
values of $N$. First, one clearly observes that, as expected from the
static results, fluctuations are larger in the canonical ensemble,
namely the system is less correlated than in the
microcanonical. Moreover, let us notice the nontrivial shape of
$C(t)$. For small $N$ one has a damped oscillatory relaxation, while,
increasing $N$, a peculiar behavior emerges: after a first stage of
relaxation, governed by a simple exponential decay, at later times a
negative bump occurs, signaling the presence of another timescale in
the system. This negative contribution to the correlation is necessary
for the vanishing of the diffusion constant: $\int_0^\infty C(t)dt$
must be zero.

%\subsection{Two timescales}

\begin{figure}[!t]
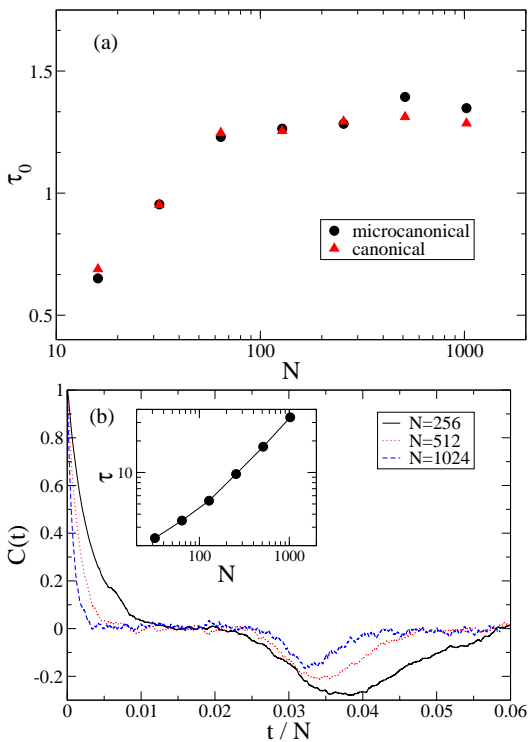

\includegraphics[width=.8\columnwidth,clip=true]{tau0.eps}
\includegraphics[width=.8\columnwidth,clip=true]{corrNlarge_resc.eps}
\caption{(color online) Panel (a): first relaxation time $\tau_0$ of the piston velocity
  correlation for different values of $N$, in the microcanonical
  (black dote) and canonical (red squares) ensemble with parameters
  $M=50$, $F=150$, $L=30$ and $k_BT=10$. Notice that the first
  relaxation saturates for large $N$, and the time $\tau_0$ reaches a
  constant value, both in the canonical and microcanonical
  ensembles. Panel (b): velocity correlation functions as a function of
  time rescaled by $N$ in the canonical ensemble with same
  parameters. In the inset the time $\tau$ shows a linear dependence
  on $N$, for large $N$.}
\label{fig5}
\end{figure}

From the above results for $C(t)$, a two-time scenario emerges. We
have the time $\tau_0$, characterizing the first exponential decay,
empirically defined as the time necessary to cross the zero axis for the first
time. In addition, we have the time $\tau(N)$ where the negative bump
occurs. The first decay of the velocity correlation function $C(t)$
saturates upon increasing the number of particles and so the time
$\tau_0$ tends to a constant value, independent of $N$ (see panel (a)
of Figure~\ref{fig5} where $\tau_0$ is plotted as a function of
$N$ in semilog scale, both for the microcanonical and the canonical
ensembles). On the other hand, we find that the second timescale
$\tau$ depends linearly on $N$, as it is shown in panel (b) of
Figure~\ref{fig5}, where $C(t)$ is plotted as a function of $t/N$. In
the inset we also plot $\tau(N)$ as a function of $N$ in log-log scale
for the canonical ensemble, showing the linear increasing with $N$
(analogous results are observed for the microcanonical ensemble).

As discussed in the next section, such a peculiar behavior, induced by
the presence of the partitioning piston, cannot be easily described by
a standard Langevin-like approach.

\section{Langevin equation}
\label{Langevin}

In the limit of $N$ and $M$ very large, the relaxation times of the
piston and of the gas particles are well separated, and one may
consider the gas particles weakly perturbed by the presence of the
piston.

Within this strong assumption, the gas distribution is fixed and
independent of the motion of the piston, and the dynamics can be
described by a master equation for the probability density function
$P(V,Y,t)$ from the velocity $V$ of the piston at position $Y$ at time
$t$. In particular, for the first moment of this distribution, it is
possible to write down the following equation (for the details refer
to the Appendix~\ref{app:B}):
%$\sigma_{y}/\Delta Y\ll 1$
% \begin{equation}
% \frac{d\left<V\right>}{dt}=\left<F(Y,V)\right>
% \end{equation}
\begin{equation}
\frac{d\left<V\right>}{dt}=\left<F_{coll}(Y,V)\right> \label{fcoll}
\end{equation}

% From Eq.~(\ref{fcoll}) it is possible to assume 
% \begin{itemize}
% \item $m \ll M$ this produce 
% \item For large $N$, $\sigma_{y}/\left< Y\right>\ll 1$, then the
% \end{itemize}

Then, the fluctuations around the equilibrium position ($Y\simeq
Y_{eq}$ and $V\simeq 0$) are described by expanding up to the first
order the right hand side of Eq.~(\ref{fcoll}), obtaining
 
%%$V\simeq \Delta V$ and $Y\simeq Y+\Delta Y$  

\begin{equation}
%\phantom{M}\frac{d Y (t)}{dt}&=& V(t)\nonumber\\
\frac{d V (t)}{dt}=-k_{N}  y-\gamma V+\xi(t), \label{eqLangevin}
\end{equation}
where the displacement $y\equiv Y-Y_{eq}$ has been introduced.  The
parameters $k_{N}$ and $\gamma$ can be calculated by means of kinetic
theory, and their explicit expressions are written in
Eq.~(\ref{explicit_expressions}) of Appendix~\ref{app:B}. One must
notice that in Eq.~(\ref{eqLangevin}) a noise term $\xi(t)$ has been
added, whose expression cannot be directly derived from the
Eq.~(\ref{fcoll}) for the mean velocity. Actually, the correlation of
the noise term can be determined by exploiting equipartition theorem
valid for equilibrium dynamics.  By requiring Maxwellian statistics
for the stationary $P(V)$, it is well known that $\xi(t)$ must be
white noise with variance
\begin{equation}
\left<\xi(t)\xi(t')\right>=2\gamma T \delta(t-t').
\end{equation}
From the linearity of Eq.~(\ref{eqLangevin}) %  it is straightforward to derive the correlation function
% \begin{equation}
% \left<v(t)v(0)\right>=\frac{T}{M}e^{-\alpha t}
% \end{equation}
it is possible to calculate the autocorrelation of velocity,
%in order to make a comparison with the numerical experiments,
obtaining:
\begin{equation}
\left<V(t)V(0)\right>=\frac{T}{M} e^{-\frac{\gamma t}{2}} \left[\cosh \left(\frac{\Delta}{2} t
\right)-\frac{\gamma \sinh \left(\frac{\Delta}{2} t  
  \right)}{\Delta}\right]  \label{correlation}
\end{equation}
where we intruduced the parameter $\Delta=\sqrt{\gamma ^2-4
  k_{N}}$, which rules the passage between underdamped and overdamped
regime. More specifically, if $\frac{Nm}{m+M}>\frac{\pi}{2}$, the system
is overdamped, else the system is underdamped.

\begin{figure}[!t]
\begin{center}
\includegraphics[width=.8\columnwidth,clip=true]{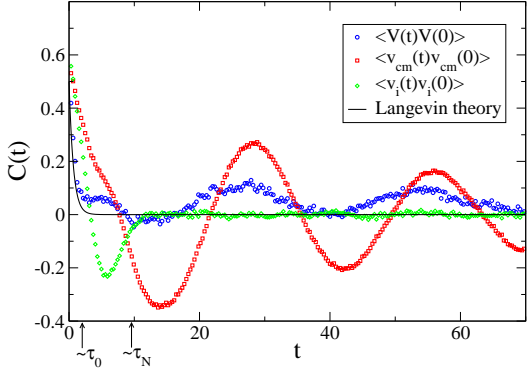}
\caption{(color online) Autocorrelation of different observables in the case of a
  piston with non-interacting particles (canonical ensemble) measured
  in numerical simulations and Langevin approximation for the piston
  velocity correlation (black line). It is possible to observe how the
  oscillation in the autocorrelation of the piston velocity
  $\left<V(t)V(0)\right>$ (blue circles) are in phase with the one of the
  center of mass of the gas particles
  $\left<v_{cm}(t)v_{cm}(0)\right>$ (red squares). With green diamonds is
  represented the autocorrelation of a single particle velocity
  $\left<v_{i}(t) v_{i}(0)\right>$. All the correlations are
  normalized to one for $t=0$. Value of the parameters: $F=150$, $T=10$,
  $M=50$, $N=500$.}
\label{fig:center_mass}
\end{center}
\end{figure}

Making a comparison between Eq.~(\ref{correlation}) and the numerical
experiments presented in Fig.~\ref{fig4}, it appears evident that the
Langevin equation is able to capture, for $N$ large, only the small time relaxation
$\tau_{0}\simeq \gamma^{-1}$, while is unable to detect the oscillation of $\langle V(t)
V(0)\rangle$, that appears for times $\tau(N)\sim N$. We report in
Fig.~\ref{fig:center_mass} the explicit comparison between the
Langevin approximation (black curve) and the piston velocity
correlation (red curve) in the non-interacting case.  The same
mismatch between analytical prediction and numerical results is
observebd also for interacting particles. The oscillations presented
by $\langle V(t) V(0)\rangle$ are related with the interplay mechanism
between the moving wall and a collective mode of the gas particles,
that make the assuption of Markovianity to fail.  We note how this
phenomenon is quite general and it is present also in the case of
non-interacting gas particles. In order to verify this point, one can
analyze a natural collective variable of the gas, i.e. the center of
mass velocity $v_{cm}(t)\equiv \frac{1}{N}\sum v_{i}(t)$. In the
simpler case of a non-interacting gas confined in a fixed volume, the
autocorrelation $\left<v_{cm}(t)v_{cm}(0)\right>$ would be trivially
equal to the one of a single particle in the gas. On the contrary this
is not true anymore with the presence of the piston, since the
different particles of the gas strongly correlates each other via the
mutual interaction piston/border. The time scale of this process is
very close to $\tau(N)$, as it can be observed in
Fig.~\ref{fig:center_mass}. Such a time scale is completely hidden if
one consider only the single particle autocorrelation
$\left<v_{i}(t)v_{i}(0)\right>$.

% } It appears evident that
% Eq.~\ref{eqLangevin} captures only the first exponential decay given
% by $\tau_{0}$

% Clearly, the stochastic term is the only one that cannot
% be derived directly from Eq.~(\ref{fcoll}).

\section{Conclusions}

In the present work we have shown, with analitical calculations in the
ideal gas case and with simulations for interacting particles, that
the fluctuations in the canonical and microcanonical ensembles~\cite{TOUCH} show
relevant differences when a partitioning object, like a moving wall,
is introduced. The relevant points that we have highlighted are the
following. First, we have shown that the interaction with the
partitioning object induces nontrivial correlations among the
particles even in the ideal gas approximation, see
Fig.~\ref{fig:center_mass} in Sec.~\ref{Langevin}, irrespectively of
the ensemble, canonical or microcanonical, where the dynamics is
studied. Then, we have shown that the Langevin approach to the
dynamics of the piston captures only partially the physics of the
system. The Langevin equation, correctly predicts only the fast time
scale, namely $\tau_0\sim \gamma^{-1}$, but fails completely to catch
the slower one, which grows linearly with the number of particles in
the partitioned system, $\tau(N)\sim N$. This second time scale is
produced by non-trivial correlations among the velocity of the gas
particles and the one of the piston which are present, quite
remarkably, also in the case of non-interacting particles, as shown in
Fig.~\ref{fig:center_mass}.

We recall that the macroscopic growth of $\tau(N)$ is related to the
particular tubular geometry of the problem, where the size of the gas
compartment is increased only in one direction. Notwithstanding the
different behavior of the largest timescale, the factor 2 of
difference between canonical and microcanonical fluctuations of the
partitioning object $\sigma_Y^2$, is independent from how the
thermodynamic limit is taken, how is clear from Eq.~(\ref{sm2}) and
Eq.~(\ref{sc2}).

We can therefore conclude that partitioning geometries with a single
\emph{macroscopic} degree of freedom which is \emph{effectively}
coupled to the motion of all the \emph{microscopic} costituents of the
system represent an eligible framework to study the dynamical
properties of small systems.

%%%%%\makeappendix
\section*{Appendix A}

\subsection{Microcanonical}
In the microcanonical ensemble at energy $E$, the invariant measure is
non-zero only on the ipersurface of constant energy $S_E$. If
$\mathcal{M}$ is a subset of $S_E$ and $d\sigma$ is the infinitesimal
surface element
\begin{equation}\label{misura microcanonica}
\mathbb{P}(x\in \mathcal{M} \subseteq S_E)=\int_{\mathcal{M}}\frac{d\sigma}{\omega(E)}\frac{1}{|\nabla \mathcal{H}|},
\end{equation} 
where $\omega(E)=\partial \Sigma(E)/\partial E$. In order to derive
the expression of the temperature of the system as a function of the
energy, we must compute $\Sigma(E)$. This quantity is given by
\begin{eqnarray}\label{sigg}
\Sigma(E)&=&\int_{\mathcal{H}<E}d^Nx\,d^Ny\,dY\,d^N\mathbf{p}\,dP\, \nonumber \\
&=&L^N\int_{\sum_i {|\mathbf{p}_i|}^2/2m + P^2/2M +FY <E}d^Ny\,dY\,d^N\mathbf{p}\,dP. \nonumber \\
\end{eqnarray}
Recalling that the volume of a $D$-dimensional sphere of radius $R$ is
$V(R)=\int_{\sum_{i}x_i^2<R^2}d^Dx=\frac{\pi^{\frac{D}{2}}}{\Gamma(\frac{D}{2}+1)}R^D$,
where $\Gamma(x)$ is the Euler Gamma, from Eq.~(\ref{sigg}) we obtain
%\begin{equation}
%\Sigma(E)=\int_{\mathcal{H}<E}d^Nx\,d^Ny\,dY\,d^N\mathbf{p}\,dP=(2m)^N\frac{\sqrt{2M}}{F}\widetilde{\Sigma}(E)
%\end{equation}
\begin{eqnarray}
\Sigma(E)&=&(2m)^N\sqrt{2M}L^N\frac{\pi^{N+\frac{1}{2}}}{\Gamma(N+\frac{3}{2})} \nonumber \\
&\times&\int_0^Y d^Ny \int_0^{E/F} dY  (E-FY)^{N+\frac{1}{2}} \nonumber \\
&=&(2m)^N\frac{\sqrt{2M}}{F}\left(\frac{L}{F}\right)^N\frac{\pi^{N+\frac{1}{2}}}{\Gamma(N+\frac{3}{2})}E^{2N+\frac{3}{2}}\nonumber \\
&\times&\int_0^1 dx\,x^N(1-x)^{N+\frac{1}{2}}
\end{eqnarray}
and, eventually, 
\begin{equation}
\Sigma(E)=(2m)^N\frac{\sqrt{2M}L^N}{F^{N+1}}\pi^{N+\frac{1}{2}}\frac{\Gamma(N+1)}{\Gamma(2N+\frac{5}{2})}E^{2N+\frac{3}{2}},
\end{equation}
and
\begin{equation}
\omega(E)=(2m)^N\frac{\sqrt{2M}L^N}{F^{N+1}}\pi^{N+\frac{1}{2}}\frac{\Gamma(N+1)}{\Gamma(2N+\frac{3}{2})}E^{2N+\frac{1}{2}}
\end{equation}
Now we can compute the temperature of the system using the
  relation $S=k_B\ln\Sigma(E)$, namely
\begin{equation}\label{temperatura micro}
k_BT=k_B\Big(\frac{\partial S}{\partial E}\Big)^{-1}=\frac{\Sigma(E)}{\omega(E)}=\frac{E}{2N+\frac{3}{2}}.
\end{equation}
Using alternative definitions of $S$, e.g. $S=k_B\ln \omega(E)$ or
$S=k_B\ln \Gamma_{\Delta E}(E)$, where $\Gamma_{\Delta E}(E)= \Sigma
(E +\Delta E) - \Sigma (E)\simeq \omega(E)\Delta E$ where $\Delta E$
is the tolerance on $E$, for $N \gg 1$ one has negligible differences
\cite{HUANG}.

We are interested in the probability density function of the position
of the piston $Y$.  Observing that for a generic phase space function
$A(\mathbf{X})$ in the microcanonical ensemble one
has~\cite{K49}:
\begin{equation}\label{marginal vera}
\rho_A(a)=\frac{1}{\omega(E)}\frac{\partial}{\partial E}\mathcal{I}(E,a),
\end{equation}
where
\begin{equation}
\mathcal{I}(E,a)=\int_{\mathcal{H}<E}\delta(A(\mathbf{x})-a)d\mathbf{x},
\end{equation}
putting $A(\mathbf{X})=Y$ one readily obtains 
\begin{eqnarray}\label{integralmarginal}
I(E,Y=\tilde{Y})&=&\int_{H<E}dY\,d^Nx\,d^Ny\,d^N\mathbf{p}_idp\,\delta(Y-\tilde{Y}) \nonumber \\ 
&=&(2m)^N\sqrt{2M}L^N\frac{\pi^{N+\frac{1}{2}}}{\Gamma(N+\frac{3}{2})}
\tilde{Y}^N(E-F\tilde{Y})^{N+\frac{1}{2}}, \nonumber \\
\end{eqnarray}
for $0<\tilde{Y}<E/F$; therefore
\begin{eqnarray}\label{prob_pos_micro}
\rho_E(Y)&=&\frac{1}{\omega(E)}\frac{\partial I}{\partial
  E}=\frac{\Gamma(2N+\frac{3}{2})}{\Gamma(N+\frac{1}{2})\Gamma(N+1)}\nonumber \\
&\times&\frac{F}{E}\Big(\frac{FY}{E}\Big)^N\Big(1-\frac{FY}{E}\Big)^{N-\frac{1}{2}}.
\end{eqnarray}
From the above result, we obtain
\begin{equation}\label{mediomicro}
\langle Y\rangle=\frac{(N+1)k_BT}{F}
\end{equation}
and
\begin{equation}
\sigma^2_Y=\frac{(N+\frac{1}{2})(N+1)}{2N+\frac{5}{2}}\Big(\frac{k_BT}{F}\Big)^2,
\end{equation}
where, in the two last equations, we used Eq.~(\ref{temperatura micro})
to express $\langle Y \rangle$ and $\sigma^2_Y$ as functions of $T$
instead of $E$.

\subsection{Canonical}
In the canonical ensemble at constant temperature $T$ with
$\beta=1/(k_BT)$, the partition function of the system is given by
\begin{eqnarray}
Z&=&\int d^Nx\,d^Ny\,dY\,d^N \mathbf{p}\,dP\,\, e^{-\beta \mathcal{H}}=\\
&=&\left(\frac{2}{\pi}\right)^{N+\frac{1}{2}}N!\,m^N\sqrt{M}\beta^{-(2N+\frac{3}{2})}F^{-(N+1)}. \nonumber 
\end{eqnarray}
We can easily compute the mean energy of the system
\begin{equation}\label{energia media canonico}
E=\langle \mathcal{H} \rangle =-\frac{\partial \ln Z}{\partial \beta}=\Big(2N +\frac{3}{2}\Big) k_BT.
\end{equation}
Now we want to find the probability distribution function of the position of the piston $Y$: starting from
\begin{equation}
\rho_\beta(Y,\{y_i\})=	\frac{e^{-\beta FY}}{\int \,dY\,d^Ny\,e^{-\beta FY}}\prod_{i}\Theta(Y-y_i),
\end{equation}
and integrating over all the $y_i$, one obtains
\begin{equation}\label{probab z canon}
\rho_\beta(Y)=\frac{Y^Ne^{-\beta FY}}{\int dY\,Y^Ne^{-\beta FY}}.
\end{equation}
The mean value of this distribution is
\begin{equation}
\langle Y \rangle =\frac{k_BT(N+1)}{F}
\end{equation}
whereas its variance is
\begin{equation}\label{fluttuazioni canon}
\sigma_Y^2=\frac{(N+1)(k_BT)^2}{F^2}.
\end{equation}

\section*{Appendix B}
 \label{app:B}

In this appendix, we detail the derivation of the Langevin equation
for the motion of the piston, following elementary kinetic theory. The
basic idea is to estimate the average force exerted by the gas
particles which collide with the piston, by calculating the average
momentum exchanged in the collisions. The following approach
  dates back to Smoluchowski~\cite{S06} and it has been used to write a Langevin
  equation for colloidal particles~\cite{DGL81}. For the variable
$y=Y-Y_{eq}$ we will derive a stochastic equation
\begin{equation}
M\frac{d^2y}{dt^2}=F_{av}(y,\dot{y})+C \eta,
\end{equation}
where $F_{av}(y,\dot{y})$ is the average force acting on the piston in
the position $Y_{eq}+y$ and velocity $\dot{y}$, $\eta$ is a white
noise and the constant $C$ can be fixed \emph{a posteriori} from the
condition $M\langle \dot{y}^2\rangle=k_BT$.

Consider the gas at equilibrium, and focus on the collision of the
piston, characterized by its mass $M$ and precollisional velocity $V$,
and a particle of the gas, which are characterized by $m$ and
$\mathbf{v}$, respectively. The collision rule is
\begin{equation}
V'=V + \frac{2m}{m+M}(v_y-V)\hspace{1 cm}v_y'=v_y - \frac{2M}{m+M}(v_y-V)
\end{equation}
where the primed quantity are postcollisional velocities, and $v_y$ is
the $y$-component of $\mathbf{v}$. The rate of such collisions can be
obtained by considering the equivalent problem of a piston, at rest,
hit by a flux of particles moving at relative velocity
$V\mathbf{\hat{y}} - \mathbf{v}$. The rate is then determined by
counting the number of point-like particles hitting the unit surface in
the infinitesimal time interval $dt$. This number corresponds to the
particles contained in a rectangle of infinitesimal base length
$\delta x$ and height $(v_y-V)\Theta(v_y-V)dt$. The step function
$\Theta(s)$ selects the condition for having a collision. Setting
$v=v_y$, the mean force exerted by the particles of the gas on the
piston is
\begin{eqnarray}
&&F_{coll}(Y,V)=\left\langle M\frac{\Delta V}{dt}\right\rangle \nonumber \\
&=&M\int_{-\infty}^\infty dv \int_0^L dx\, \rho(x,Y-r_0') \nonumber \\ 
&\times&\phi(v) (V'-V)(v-V)\Theta(v-V)\nonumber \\
&=&\frac{2mM}{m+M}\nonumber \\
&\times&\int_{-\infty}^\infty dv \int_0^L dx\, \rho(x,Y-r_0')\phi(v)\Theta(v-V) (v-V)^2
\nonumber \\
\end{eqnarray}
where $\phi(v)$ is the equilibrium distribution of velocities of the
gas, i.e. $\phi(v)=\sqrt{\frac{m}{2\pi k_BT}}e^{-\frac{mv^2}{2k_BT}}$
and $\rho(x,Y)$ is the spatial density of particles in the proximity
of the piston. At equilibrium, this density is uniform on all the
available volume and, therefore, depends on the position of the piston
$Y$. Carrying on the integration on the spatial coordinates, we obtain
\begin{equation}\label{integrale urti}
F_{coll}(Y,V)=\frac{2mM}{m+M}\lambda\int_V^\infty\,dv(v-V)^2\phi(v) 
\end{equation}
where $\lambda=\frac{N}{Y}$. We note that the equilibrium properties
of the gas used in the derivation of this equation don't depend on the
choice of the ensemble. Of course, $F_{av}(y,\dot{y})$ is nothing but
$F_{coll}-F$.

In order to decouple the motion of the piston from the one of the gas
molecules it's necessary to assume that $M\gg m$ and that, moreover,
$V$ is always small if compared to the thermal velocity of the
particles $v_m=\sqrt{\frac{2k_BT}{m}}$: the expansion of the integral
in Eq.~(\ref{integrale urti}) in powers of $\sqrt{\frac{m}{M}}$, will
give the viscous drag force appearing in the Langevin equation of
motion.  Defining $g=\sqrt{\frac{m}{2k_BT}}(v-V)$ and expanding
perturbatively $\phi(v)$ as a function of $g$
\begin{eqnarray}
e^{-\frac{m}{2k_BT}v^2}&=&e^{-\left(g+\sqrt{\frac{m}{2k_BT}}V\right)^2}\simeq
e^{-g^2-\sqrt{\frac{2m}{k_BT}}gV} \nonumber \\
&\simeq&
e^{-g^2}\left(1-\sqrt{\frac{2m}{k_BT}}gV\right)
\end{eqnarray}
we can compute the integral, performing the change of variables $v\to
g$
\begin{equation}
\frac{2k_BT}{m\sqrt{\pi}}\int_0^\infty g^2
e^{-g^2}\left(1-\sqrt{\frac{2m}{k_BT}}gV\right)dg=\frac{k_BT}{2m}-\sqrt{\frac{2k_BT}{\pi
    m}}V \nonumber
\end{equation}
namely
\begin{equation}
F_{coll}=\frac{N}{Y}\left[ \frac{M}{m+M} k_BT - 2\frac{M}{m+M}\sqrt{\frac{2mk_BT}{\pi}}V\right]
\end{equation}
Expanding the previous expression at the first order in $y$ and $V$
around the equilibrium position of the piston $Y_{eq}$, defined by the
condition $F=F_{coll}$ and $V=0$, we obtain a linear Langevin
equation. The equilibrium conditions are
\begin{equation}
\frac{M}{m+M}k_BT\frac{N}{Y_{eq}}=F \;\; \textrm{and} \;\; V_{eq}=0
\end{equation} 
and therefore 
\begin{equation}
Y_{eq}=\frac{NMk_BT}{F(m+M)}.
\end{equation}
The Langevin equation has the shape 
\begin{equation}
\frac{d^2y}{dt^2}=-k_N y-\gamma v+\frac{C}{M} \eta,
\label{lastlang}
\end{equation}
where
\begin{equation}
\gamma=\frac{2F}{M}\sqrt{\frac{2m}{\pi k_BT}} \;\; \textrm{and} \;\;
k_N=\frac{F^2(m+M)}{M^2Nk_BT}.\label{explicit_expressions}
\end{equation}
It is easy to compute the correlation function
\begin{equation}
\langle V(t)V(0) \rangle = \frac{k_BT}{M}e^{-\frac{\gamma}{2}}
\left[ \cosh\left(\frac{\Delta}{2}t\right)-\frac{\gamma}{\Delta}\sinh\left(\frac{\Delta}{2}t\right)\right],
\end{equation}
where $\Delta=\sqrt{\gamma^2-4k_N}$. Let us note that for any finite
$N$ (i.e. $k_N\ne 0$) one has $\int_0^\infty\langle V(t)V(0) \rangle
dt=0$.

\begin{acknowledgments}
We thank M. Falcioni and A. Puglisi  for useful discussions. The work
of AS is supported by the Granular Chaos project, funded by the
Italian MIUR under the grant number RIBD08Z9JE.
\end{acknowledgments}

%\section*{References}

\end{document}